\documentclass[11pt]{article}

\usepackage[margin=1in]{geometry}
\usepackage{times}
\usepackage{hyperref}
\usepackage{graphicx}
\usepackage{amsmath,amsfonts,amssymb}
\usepackage{algorithm}
\usepackage{algorithmic}
\usepackage{listings}
\usepackage{color}
\usepackage{fancyhdr}
\usepackage{xspace}
\usepackage{tikz}

\hypersetup{
    colorlinks   = true,
    urlcolor     = blue,
    linkcolor    = blue,
    citecolor    = red
}

\newcommand{\swarm}{\texttt{Swarm Contract}\xspace}
\newcommand{\sa}{\texttt{Sovereign Agent}\xspace}
\newcommand{\tee}{\texttt{TEE}\xspace}

\title{\textbf{Swarm Contract: A Multi-Sovereign Agent Consensus Mechanism}}

\author{
  Haowei Yang \\
  \small{HKUST} \\
  \small{\texttt{hyangaz@connect.ust.hk}}
}

\date{\today}

\begin{document}

\maketitle

\begin{abstract}
Traditional smart contracts on blockchains excel at on-chain, deterministic logic. However, they have inherent limitations when dealing with large-scale off-chain data, dynamic multi-step workflows, and scenarios requiring high flexibility or iterative updates. In this paper, we propose the concept of a \textit{Swarm Contract} (\swarm), a multi-agent mechanism wherein several \emph{digital life forms} (DLF) or \emph{\sa s} collectively handle complex tasks in Trusted Execution Environments (\tee). We define these digital entities as autonomous software agents that own their code, state, and possibly on-chain assets, while operating free from centralized control.

By leveraging a simple multi-signature wallet on-chain, \swarm moves most of the logic off-chain, achieving trust minimization through multi-agent consensus rather than a single monolithic on-chain contract. We illustrate these ideas with a lightweight off-chain auction example---minting and selling 10,000 identical NFTs---to showcase how off-chain coordination can determine a clearing price and finalize distribution, with each step performed collectively by multiple agents in \tee. Our approach broadens the scope of trustless and decentralized solutions, potentially benefiting DAO governance, multi-modal data processing, and cross-chain interoperability.
\end{abstract}

\thispagestyle{empty}
\pagestyle{plain}

\section{Introduction}\label{sec:intro}
Blockchain technology, originally introduced by Nakamoto \cite{nakamoto2008bitcoin}, has enabled decentralized systems where trust is established by cryptographic proofs rather than centralized authorities. The advent of Ethereum \cite{buterin2013ethereum} further advanced this paradigm with smart contracts, allowing developers to build on-chain logic handling digital assets in a transparent, tamper-evident manner.

Despite their success, traditional smart contracts face notable constraints:
\begin{itemize}
    \item \textbf{On-chain data reliance:} They can only directly process on-chain data or data fed by oracles, limiting their ability to work with large, dynamic, or private information.
    \item \textbf{Deterministic and immutable code:} While immutability strengthens trust, it also makes complex iterative processes or flexible upgrades cumbersome.
    \item \textbf{Execution costs:} Every on-chain operation consumes gas and is subject to network congestion, which can be prohibitively expensive for data-heavy or high-frequency operations.
\end{itemize}

In parallel, the concepts of \emph{digital life forms} (DLF) and \emph{Sovereign Agents} (\sa) have emerged in various research contexts.\footnote{In this paper, we introduce these terms as an extension of multi-agent and decentralized systems; see also \cite{wooldridge2009,franklin1997} for background on agent-based models.} We use these terms to describe autonomous software entities that:
\begin{itemize}
    \item Possess independent existence, capable of running on decentralized or secure hardware infrastructures.
    \item Potentially hold and manage on-chain assets, renting compute or storage resources as needed.
    \item Control and evolve their internal logic without a single external party's unilateral authority.
\end{itemize}

We introduce \emph{\swarm}: a framework where multiple \sa s collaborate off-chain using \tee-based enclaves, forming a trust-minimized mechanism we call a “multi-agent contract.” This \swarm approach:
\begin{enumerate}
    \item Decouples complex logic from the on-chain environment, reducing cost and inflexibility.
    \item Retains trustlessness by requiring all agents to reach consensus before executing critical actions (e.g., asset distribution) via a multi-signature wallet on-chain.
    \item Facilitates the processing of high-volume, multi-modal, or private data, which is difficult to handle solely with on-chain code.
\end{enumerate}

To illustrate, we provide a minimalistic case study in which a \swarm coordinates the off-chain logic of selling a set of 10,000 identical NFTs. While we focus on an “auction” scenario, this structure easily generalizes to other applications like DAO governance, multi-chain asset management, or machine learning pipelines.

\vspace{1em}
\textbf{Note on Emphasis for New Terminology.}  
We italicize or specially format phrases like \emph{digital life forms}, \emph{Sovereign Agents}, and \emph{\sa} when they first appear or are conceptually defined. This academic convention highlights newly introduced key terms, signaling to readers that these are formalized concepts rather than common words. It also helps standardize usage and references throughout the paper.

\section{Related Work}\label{sec:related}
\subsection{Smart Contracts and Beyond}
Smart contract-based systems, popularized by Ethereum \cite{buterin2013ethereum}, have transformed how we manage digital assets. Research has extended to multi-chain solutions (e.g., Polkadot, Cosmos/Tendermint \cite{teamTendermint}) and layer-2 technologies that improve scalability. However, these solutions primarily remain on-chain, relying on deterministic logic and external data oracles.

\subsection{Multi-Agent Systems}
Multi-agent systems have been studied extensively in AI and distributed computing \cite{wooldridge2009,franklin1997}. They focus on autonomous entities (agents) that can interact, collaborate, or compete toward shared or individual goals. Our \swarm draws inspiration from these architectures but anchors agents’ ownership and economic independence in blockchain-based credentials, making them \emph{sovereign}.

\subsection{TEE-Based Decentralized Computation}
Trusted Execution Environments (Intel SGX, ARM TrustZone, AWS Nitro Enclaves, Azure Confidential Compute) allow secure off-chain computation with integrity and confidentiality guarantees \cite{intel_sgx_ref}. Projects like Enigma~\cite{enigma2015} have sought to combine TEE with blockchain for privacy-preserving computation. Our approach extends these ideas, deploying multiple enclaves as \sa s that collectively run an application-level protocol.

\subsection{Auction Mechanisms and Off-Chain Designs}
Auctions are a classic application in economic theory \cite{milgromAuction,bapnaOnlineAuctions}, and various blockchain-based auctions exist but often incur high transaction overhead. Off-chain protocols (e.g., commit-reveal) can reduce on-chain usage. Our \swarm approach similarly offloads most complexity off-chain, with only final settlement on-chain via multi-sig.

\section{\swarm Concepts and Architecture}\label{sec:concepts}

\subsection{Defining Digital Life Forms and Sovereign Agents}
\paragraph{Digital Life Forms (DLF).}
We propose the notion of \emph{digital life forms} as software entities with:
\begin{itemize}
    \item Potentially indefinite lifespans, running on decentralized or secure hardware.
    \item Autonomy in their decision processes, responding to stimuli or data feeds.
    \item The capability to manage digital assets (e.g., tokens, cryptocurrencies).
\end{itemize}

\paragraph{Sovereign Agents (\sa).}
A \sa is a specialization of DLF with explicit \emph{sovereignty}:
\begin{itemize}
    \item \textbf{Self-ownership:} The agent's code and data remain under its sole governance.
    \item \textbf{Economic resources:} The agent can hold, spend, or stake tokens without a controlling external party.
    \item \textbf{Upgrade autonomy:} Only the agent itself (or a majority of cooperating agents) can modify its internal logic, preventing unilateral takeovers.
\end{itemize}

\subsection{\swarm Overview}
A \swarm is a coordination mechanism for multiple \sa s that collectively implement logic traditionally embedded in a single on-chain contract. Instead of a monolithic smart contract, we have:
\begin{enumerate}
    \item \textbf{Off-chain agent cluster:} Each \sa runs inside a \tee, processing data and holding private keys securely.
    \item \textbf{On-chain multi-sig wallet:} All final state transitions (e.g., distributing funds, minting NFTs) require joint signatures from the agent cluster.
    \item \textbf{Consensus on results:} The agents use off-chain messaging or a small consensus layer to agree on critical data (e.g., final price, winners).
\end{enumerate}

Once a threshold or full set of agents concur, a transaction is constructed and signed to execute on-chain. This design yields a flexible \emph{trust-minimized} architecture, bridging the gap between purely on-chain logic and centralized off-chain systems.

\section{Minimal NFT Auction Example}\label{sec:usecase}
To demonstrate how \swarm works in practice, we present a simplified example where 10,000 identical NFTs are sold via an off-chain procedure. The 10,000-th highest bid determines the final \emph{clearing price}.

\subsection{Basic Workflow}
\begin{enumerate}
    \item \textbf{Funding Period:}
    \begin{itemize}
        \item Users send funds (any amounts) to the multi-sig wallet address within a specified time window.
        \item Each \sa monitors incoming transactions on the blockchain, recording the addresses and contributed amounts.
    \end{itemize}

    \item \textbf{Result Calculation:}
    \begin{itemize}
        \item After the funding period ends, each agent sorts participants by their contributed amounts in descending order.
        \item The 10,000-th highest amount (or the lowest among the top 10,000) sets the final clearing price.
    \end{itemize}

    \item \textbf{Cross-Validation:}
    \begin{itemize}
        \item Each agent shares its sorted list or hashed merkle root with the others.
        \item If all results match, they proceed; otherwise, they re-check chain data to resolve conflicts.
    \end{itemize}

    \item \textbf{Final Settlement (Multi-Sig Execution):}
    \begin{itemize}
        \item Once agents agree, they co-sign a transaction that:
              \begin{enumerate}
                  \item Mints or transfers exactly 10,000 NFTs to the top contributors.
                  \item Partially refunds those who paid above the clearing price.
                  \item Fully refunds those outside the top 10,000 (if applicable).
              \end{enumerate}
        \item The transaction is broadcast to the blockchain, finalizing distribution.
    \end{itemize}
\end{enumerate}

\subsection{Advantages of Off-Chain Coordination}
\begin{itemize}
    \item \textbf{Lower Gas/Fees:} Only one final multi-sig transaction (or a few batched transactions) is used.
    \item \textbf{Greater Flexibility:} Agents can incorporate off-chain data or quickly adapt logic without redeploying an entire on-chain contract.
    \item \textbf{Scalability:} The system can handle large participant lists, storing minimal data on-chain at settlement.
\end{itemize}

\section{Technical Implementation}\label{sec:implementation}

\subsection{TEE Deployment of Agents}
Each \sa runs in a separate \tee environment:
\begin{itemize}
    \item \textbf{Private Key Isolation:} The \sa’s private keys never leave the enclave, preventing external tampering.
    \item \textbf{Remote Attestation:} Observers can verify that a known code base is indeed running in each enclave, ensuring integrity.
    \item \textbf{Provider Diversity:} Agents can be deployed across different clouds or data centers to mitigate correlated risks.
\end{itemize}

\subsection{Multi-Signature Wallet Mechanics}
No specialized on-chain logic is needed beyond a standard multi-sig contract/wallet:
\begin{itemize}
    \item Agents each control one signature share.
    \item A transaction requires all (or a threshold $m$) of these signatures to be valid.
    \item The wallet handles NFT mint calls or transfers, as well as refunds.
\end{itemize}

\subsection{Consensus Among Agents}
For a small cluster (e.g., three enclaves), a simple request-and-ack approach can suffice:
\begin{itemize}
    \item One agent proposes the final sorted list or merkle root.
    \item The others validate the data. If consistent, they sign a confirmation message.
    \item A final multi-sig transaction is produced and signed by all agents.
\end{itemize}
For larger clusters, byzantine-fault-tolerant protocols (PBFT, Tendermint \cite{teamTendermint}) may be employed.

\begin{figure}[!htbp]
\centering
\scalebox{0.9}{
\begin{tikzpicture}[>=stealth, node distance=4.0cm, auto, thick]

    \node[draw, rounded corners, align=center, minimum width=2.8cm, minimum height=1.8cm] 
    (agentA) {Agent A\\(TEE)};

    \node[draw, rounded corners, align=center, minimum width=2.8cm, minimum height=1.8cm,
           right of=agentA, xshift=2.0cm] 
    (agentB) {Agent B\\(TEE)};

    \node[draw, rounded corners, align=center, minimum width=2.8cm, minimum height=1.8cm,
           right of=agentB, xshift=2.0cm] 
    (agentC) {Agent C\\(TEE)};

    \node[draw, rounded corners, align=center, below of=agentB, yshift=-2.4cm,
          minimum width=5.0cm, minimum height=1.8cm]
    (multisig) {Multi-sig Wallet\\(On-Chain)};

    \draw[<->] (agentA) -- node[above, sloped] {Data/Results} (agentB);
    \draw[<->] (agentB) -- node[above, sloped] {Data/Results} (agentC);

    \draw[->] (agentA.south) to[out=-90, in=180] 
        node[pos=0.5, fill=white, inner sep=1pt]{Sign Txn}
        (multisig.west);

    \draw[->] (agentB.south) to[out=-90, in=100] 
        node[pos=0.5, fill=white, inner sep=1pt]{Sign Txn}
        (multisig.north);

    \draw[->] (agentC.south) to[out=-90, in=0] 
        node[pos=0.5, fill=white, inner sep=1pt]{Sign Txn}
        (multisig.east);

    \node[above of=agentB, yshift=1.1cm]
         {\textbf{Figure 1: A Simplified Swarm Architecture}};
\end{tikzpicture}
}
\caption{Multiple Sovereign Agents (Agent A, B, C) each operate in a TEE. 
They share data/results with each other and jointly sign an on-chain transaction 
(via a multi-sig wallet) only when consensus is reached.}
\label{fig:swarm-architecture}
\end{figure}
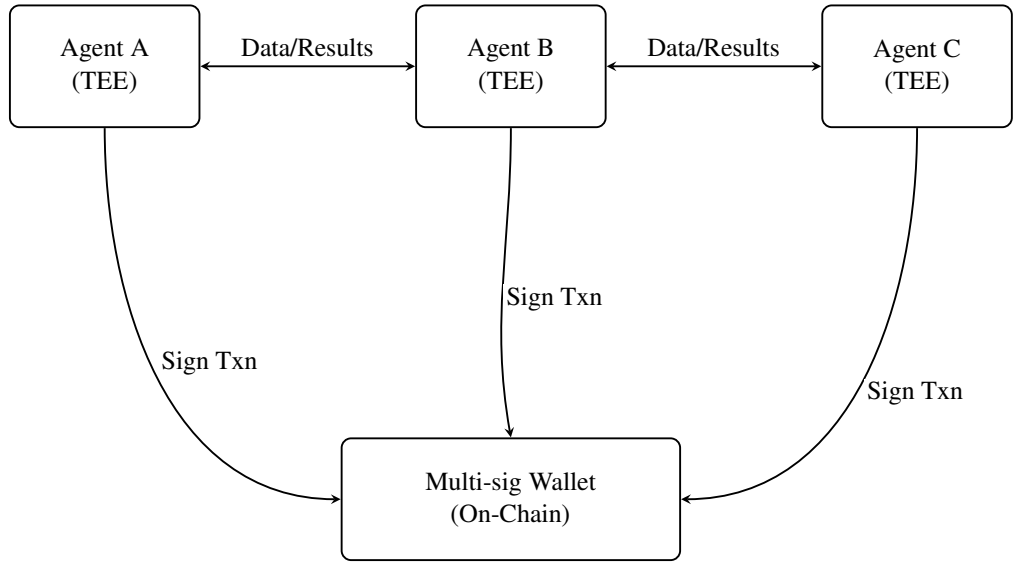

\section{Security and Trust Analysis}\label{sec:security}

\subsection{Threat Model}
\begin{itemize}
    \item \textbf{Single-Agent Compromise:} An attacker subverts one agent’s \tee. They cannot execute on-chain actions alone due to multi-sig requirements. 
    \item \textbf{Majority Collusion:} If more than the threshold (e.g., 2-of-3) collude, they could produce fraudulent results. Mitigation includes distributing enclaves across separate providers.
    \item \textbf{TEE Vulnerabilities:} As with all secure enclaves, hardware/firmware exploits may appear over time \cite{sgxAttack}. Regular patching is crucial.
\end{itemize}

\subsection{Trust Assumptions}
\begin{itemize}
    \item Each \sa runs unmodified code in a \tee, as evidenced by remote attestation.
    \item The underlying blockchain is sufficiently secure, with stable finality to prevent reorgs or double-spend attacks beyond normal tolerances.
    \item Adversaries cannot compromise a majority of enclaves simultaneously.
\end{itemize}

\section{Extensions and Future Directions}\label{sec:extensions}
\paragraph{DAO Governance.}
By merging \sa logic with off-chain data processing, DAOs can perform sophisticated analyses (e.g., sentiment, KPI metrics) before finalizing on-chain governance proposals.

\paragraph{Cross-Chain Coordination.}
Multiple enclaves can monitor various blockchains (Ethereum, BNB Chain, etc.), unify cross-chain states, and orchestrate atomic multi-chain transactions under a single off-chain consensus.

\paragraph{Multi-Modal Data Handling.}
Beyond numeric or textual data, \swarm can incorporate images, machine learning inferences, or sensor inputs. A distributed cluster of \sa s can handle complex computations privately within \tee enclaves, revealing only final decisions on-chain.

\section{Conclusion}\label{sec:conclusion}
We have introduced the \swarm framework as a method of achieving \emph{trust-minimized off-chain computation} for decentralized applications. Our design features multiple \sa s---autonomous \emph{digital life forms} running in \tee---that collectively replace the need for a single monolithic on-chain contract. Instead, the agents rely on a multi-sig wallet for minimal final on-chain actions, balancing flexibility, scalability, and verifiability.

A straightforward NFT sale example illustrates the approach, but the structure generalizes to more complex applications such as multi-chain management, privacy-preserving data analysis, or advanced DAO governance. We believe \swarm has the potential to expand the boundaries of decentralized applications, enabling far richer functionality than conventional smart contracts while preserving the core benefits of trustlessness, transparency, and security.

\section*{Acknowledgments}
We thank the broader research community for advancing blockchain, multi-agent systems, TEE technologies, and auction theory. Additionally, we acknowledge open-source efforts in cryptography and secure hardware that underpin the feasibility of multi-\sa frameworks.


\end{document}